\documentclass[12pt,preprint]{aastex}
\shorttitle{OGLE-2016-BLG-0693LB}

\shortauthors{Ryu et al.}
\begin{document}
\title{OGLE-2016-BLG-0693LB: Probing the Brown Dwarf Desert with Microlensing}
\author{\textsc{Y.-H. Ryu$^{1}$, A. Udalski$^{2}$, J. C. Yee$^{3}$ \and M. D. Albrow$^{4}$, S.-J.
Chung$^{1,8}$, A. Gould$^{1,5,6}$, C. Han$^{7}$, K.-H. Hwang$^{1}$,
Y. K. Jung$^{3}$, I.-G. Shin$^{3}$, W. Zhu$^{6}$, S.-M. Cha$^{1,9}$,
D.-J. Kim$^{1}$, H.-W. Kim$^{1}$, S.-L. Kim$^{1,8}$, C.-U.
Lee$^{1,8}$, Y. Lee$^{1,9}$, B.-G. Park$^{1,8}$,
R. W. Pogge$^{6}$ \\
(The KMTNet Collaboration)\\
P. Pietrukowicz$^{2}$, S. Koz{\l}owski$^{2}$, R. Poleski$^{2,6}$, J.
Skowron$^{2}$, P. Mr\'{o}z$^{2}$, M. K. Szyma\'{n}ski$^{2}$, I.
Soszy\'{n}ski$^{2}$, M. Pawlak$^{2}$, K. Ulaczyk$^{2}$\\(The OGLE
Collaboration)} }

\affil{$^{1}$Korea Astronomy and Space Science Institute, Daejon
34055, Korea}

\affil{$^{2}$Warsaw University Observatory, Al. Ujazdowskie 4,
00-478 Warszawa, Poland}

\affil{$^{3}$Smithsonian Astrophysical Observatory, 60 Garden St.,
Cambridge, MA 02138, USA}

\affil{$^{4}$University of Canterbury, Department of Physics and
Astronomy, Private Bag 4800, Christchurch 8020, New Zealand}

\affil{$^{5}$Max-Planck-Institute for Astronomy, K\"{o}nigstuhl 17,
69117 Heidelberg, Germany}

\affil{$^{6}$Department of Astronomy, Ohio State University, 140 W.
18th Ave., Columbus, OH 43210, USA}

\affil{$^{7}$Department of Physics, Chungbuk National University,
Cheongju 28644, Republic of Korea}

\affil{$^{8}$Astronomy and Space Science Major, Korea University of
Science and Technology, Daejeon 34113, Korea}

\affil{$^{9}$School of Space Research, Kyung Hee University, Yongin,
Kyeonggi 17104, Korea}

\begin{abstract}
We present an analysis of microlensing event OGLE-2016-BLG-0693,
based on the survey-only microlensing observations by the OGLE and
KMTNet groups. In order to analyze the light curve, we consider the
effects of parallax, orbital motion, and baseline slope, and also
refine the result using a Galactic model prior. From the
microlensing analysis, we find that the event is a binary composed
of a low-mass brown dwarf ($49^{+20}_{-18}\,M_J$) companion and a K-
or G-dwarf host, which lies at a distance $5.0\pm0.6$ kpc toward the
Galactic bulge. The projected separation between the brown dwarf and
its host star is less than $\sim 5$ AU, and thus it is likely that
the brown dwarf companion is located in the brown dwarf desert.
\end{abstract}

\section{INTRODUCTION}
Brown dwarfs are generally known as failed stars that lack
sufficient mass to ignite hydrogen burning. The upper mass limit of
brown dwarfs so defined therefore varies from $0.07\,M_{\sun}$ to
$0.09\,M_{\sun}$ according to the metallicity. On the other hand,
the lower mass limit is defined by the onset of deuterium burning,
i.e., about $13\,M_J$ \citep{bur01}. Due to their low luminosity,
brown dwarfs can be hard to detect, particularly when distant from
Earth. By contrast, the microlensing technique has an advantage to
detect brown dwarfs at great distances, up to the Galactic bulge,
because microlensing events can be detected regardless of the
brightness of the lens system.

The discoveries of brown dwarfs can provide essential information on
formation theory between giant planets and stars. Presently, 19
brown dwarfs have been detected by the microlensing method. Although
the sample of microlensing brown dwarfs is still a small, it
includes various types of brown dwarf systems such as isolated brown
dwarfs \citep{poi05,gou09,zhu16,chu17}, brown dwarf companions to
dwarf stars
\citep{bac12,boz12,shi12a,shi12b,par13,par15,str13,jun15,ran15,shv16a},
a brown dwarf binary \citep{cho13}, and a brown dwarf hosting a
planet \citep{han13,shv17}.

Microlensing is particularly sensitive to detecting brown dwarfs
around low-mass stars. Recently, \citet{shv16b} presented
statistical results from `second-generation' microlensing surveys by
the Optical Gravitational Lensing Experiment \citep[OGLE;][]{uda15},
Microlensing Observations in Astrophysics \citep[MOA;][]{bon01}, and
Wise microlensing groups. They analyzed 224 microlensing events
observed by all three teams during four observation-seasons, and
suggested that the frequency of brown dwarf companions is relatively
high in microlensing events. In addition, microlensing experiments
are entering a new era, with surveys having wider fields of view and
a higher cadence such as the Korea Microlensing Telescope Network
\citep[KMTNet;][]{kim16}. The KMTNet project was designed for the
purpose of round-the-clock observing using three 1.6 m wide field
optical telescopes located in Chile, South Africa, and Australia.
Hence, all microlensing events are monitored at high cadence, making
it possible densely monitor perturbations generated by brown dwarfs.

Due to the lack of discovery of brown dwarfs with respect to that of
planets, there is a gap in the distribution between two populations
of planets and stellar binaries that the so-called ``brown dwarf
desert". The brown dwarf desert exists over the mass-ratio range of
$0.01<q<0.1$ \citep{hal00,cha03}. \citet{mar00} defined the brown
dwarf desert as the paucity of brown dwarf companions to solar-type
stars at separations of $<$ 3 AU. \citet{gre06} quantified this
desert as the intersection between two diverging power law mass
functions, one for planets and one for stellar companions, which
both decline toward brown dwarf masses. They therefore argued that
the brown dwarf desert represents incontrovertible evidence for
different formation mechanisms for stars and planets. However, the
limits of the brown dwarf desert in separation and as a function of
host mass are currently set by observational biases. Uncovering the
true nature of this desert and its boundaries provides crucial
information for understanding different formation mechanisms for
stars and planets.

The microlensing method is optimal for probing the nature of the
brown dwarf desert at intermediate separations (1 -- 10 AU) and
exploring its outer edge, if one exists. \citet{shv16b} suggest that
the brown dwarf desert might actually reach a minimum at
Super-Jupiter masses for their microlensing sample, which probes
companions to M dwarfs at wider separations than RV studies.
However, they state that more data are needed to confirm and
characterize this minimum. The stellar companion mass function in
this region is well understood \citep{duq91,rag10}, but the
lower-mass bound of the desert has not been determined. Microlensing
can complete our understanding of the brown dwarf desert by
measuring the mass function of companions in this region into the
planetary regime \citep{gou10,cas12,suz16}. As with other
techniques, microlensing is even more sensitive to brown dwarfs than
to planets, and \citet{gau02} suggested that $\sim 25\,\%$ of brown
dwarf companions from 1 -- 10 AU should be detectable in
microlensing events if they exist. To date, planets have been the
focus of microlensing, but a full analysis of all 2-body lens events
will reveal the true extent of the brown dwarf desert and its
implications for star and planet formation.

In this paper we report the discovery of a brown dwarf located in
the brown dwarf desert from data taken by the OGLE and KMTNet survey
experiments. The structure of this paper is as follows. In Section
2, we describe the discovery and survey observations, and present
the light curve modeling in Section 3. In Section 4, we derive the
physical properties of the lens system. In Section 5, we finish with
a discussion of our findings.

\section{OBSERVATION}
OGLE-2016-BLG-0693 was announced as a new microlensing event by the
OGLE Early Warning System (EWS) on 2016 Apr 15 UT 20:12
\citep{uda94,uda03,uda15}, and occurred on a star located toward the
Galactic bulge with equatorial coordinates
$(\alpha,\delta)_{\rm{J2000}}$=($17^{\rm{h}}49^{\rm{m}}55\fs29,$
$-23\arcdeg 2\arcmin 23\farcs8)$ corresponding to Galactic
coordinates $(l,b)=(5\fdg548,2\fdg220)$. OGLE observed using its 1.3
m Warsaw Telescope with a 1.4 deg$^2$ camera at the Las Campanas
Observatory in Chile. This event was also observed by KMTNet and
MiNDSTEp. In this paper, we used data based on the survey-only
microlensing observations of the OGLE and KMTNet groups. KMTNet
independently observed the event using three 1.6 m telescopes with 4
deg$^2$ camera in Chile (KMTC), South Africa (KMTS), and Australia
(KMTA) \citep{kim16}.

In this study, we used the 491 OGLE data points taken over seven
years, and the data of KMTC, KMTS, and KMTA (98, 124, and 59 points,
respectively). Note that we removed several outlier data points for
each data set. We re-scaled the photometry errors of the individual
data sets in order to make $\chi^2$ per degree of freedom unity for
the best-fit model. In the case of the KMTNet data, the event is
located in a newly added field (BLG20) with the observation cadence
of $\simeq0.4$ ${\rm hr}^{-1}$ during 2016, and thus there are
relatively few data points up to our adopted cut off of
$\textrm{HJD}^\prime(\textrm{HJD}-2450000)\sim7568$. Despite the low
observation cadence, the caustic entrance was detected by KMTS. We
present the light curve of the event in Figure 1. The event has a
caustic crossing feature of U-shape that is typical of binary
lensing. Unfortunately, the caustic exit was not observed due to the
weather conditions.

\section{BINARY LENS MODELING}
The slope in the baseline data is clearly apparent in Figure 1. This
cannot be intrinsic to the source because it would imply a period of
several decades, which would be completely inconsistent with the
properties of such dim stars. On the other hand, such long term
trends are a well known artifact induced by proper motions of
neighboring bright stars. Therefore we include a linear trend in the
blended flux to account for this effect.

In order to find all possible lensing models that can explain the
light curve of the event, we first conduct a grid search for $(s, q,
\alpha)$ parameters, where $s$ is the projected separation between
the companion and its host star, $q$ is the mass ratio of the two
components of the lens system, and $\alpha$ is the angle of the
source trajectory. In addition, the four lensing parameters
$t_0,u_0,t_{\rm E}$, and $\rho$ are needed to fit a binary-lens
light curve: $t_0$ and $u_0$ are respectively the time and
separation of the source star's approach to a reference position
(for which we adopt the center of mass), $t_{\rm E}$ is the Einstein
ring crossing time corresponding to the total mass of the lens
system, and $\rho$ is the source radius normalized to the angular
Einstein radius ($\theta_{\rm E}$). For a grid values of $(s, q,
\alpha)$ parameters, we find the best-fitting lensing model by
optimizing the other parameters using the Markov Chain Monte Carlo
(MCMC) approach. For the case of $\alpha$, we seed it on a grid of
six equally spaced starting values but then allow it to vary as a
chain parameter to reduce calculation time. Since the major purpose
of the grid search is to find crude solutions used for a starting
point, we use a fixed value for the baseline slope of 0.03 ${\rm
yr^{-1}}$.

In Figure 2, we present the $\Delta\chi^2$ map for $(s,q)$
parameters, and identify two local solutions at close $(s<1)$ and
wide $(s>1)$ separations, which is the well-known close/wide binary
degeneracy \citep{dom99}. In the MCMC fitting process, we calculate
the finite-source lensing magnification using the inverse
ray-shooting method when the source passes the caustic
\citep{kay86,sch88,wam97}, while outside of the caustic, the
magnification is calculated using the semi-analytic hexadecapole
approximation \citep{pej09,gou08}. We also consider the
limb-darkening effect by the source star. On the surface of the
source star, the brightness profile is modeled by
\begin{equation}
I_{\lambda}(\phi)=\frac{F_0}{\pi\theta_*^2}\left[1-\Gamma_{\lambda}(1-\frac{3}{2}\cos\phi)\right],
\end{equation}
where $F_0$ is the total flux of the source star, $\phi$ is the
angle of the line of sight relative to the normal to the star's
surface, and $\Gamma$ is the linear limb-darkening coefficient
\citep{alb99}. The limb-darkening coefficient $(\Gamma_{\lambda})$
is related to the linear limb-darkening coefficient $u$ of
\citet{cla00} as
\begin{equation}
\Gamma_{\lambda}=\frac{2u}{3-u}.
\end{equation}
We adopt $\Gamma_{\lambda}=0.54$ according to the source type (see
Section 4 for more details).

For each local minimum of the grid search, we perform an
optimization using the MCMC fitting process by allowing the eight
parameters including the slope to vary. In the case of long
timescale events, since the position of observer is changed due to
Earth's orbital motion during the lensing phenomenon, the light
curve is changed due to this parallax effect. Therefore we include
the parallax effect in our model, and perform the fitting of the
light curve by adding two parallax parameters $\pi_{{\rm{E}},N}$ and
$\pi_{{\rm{E}},E}$ to the standard models with the slope
\citep{gou92,gou04}.

In Figure 3. we present the light curves and caustic structures for
the best-fit parallax models at close ($s<1$) and wide ($s>1$)
separations. The best fitting parameters and their $1\,\sigma$
uncertainties are shown in Table 1, where the error of each
individual parameter is the standard deviation determined from the
distribution of MCMC chains. As shown in Table 1, the lens system
has a small mass ratio $(\sim 0.02)$, and has an unusually long
timescale $t_{\rm E}$ and uncertainty compared to typical
microlensing events.

Figure 4 shows that both parallax parameters are consistent with
zero at the $2\,\sigma$ level. It had been common practice in the
past to suppress the parallax parameters in such cases on the
grounds that the parallax was ``not measured" to be different from
zero. However, this viewpoint was not correct, as discussed in the
Appendix of \citet{han16}. The key point is that $\pi_{\rm E}$ is
known a priori to be different from zero. so the fact that it is
shown to be consistent with zero to within small errors is important
information. In the present case, this parallax constraint will
constrain the lens system not to be very close (and low mass). This
will be combined with a flux constraint to provide reasonably strong
constraints on all parameters. Hence this ``non-measurement"
actually plays an important role.

The question of whether to introduce orbital motion parameters is
governed by the same fundamental issues of physics as for parallax
parameters, but falls into a very different regime and therefore
leads to a substantially different treatment. Just as we know a
priori that all microlenses have finite (non zero) microlens
parallax, we also know that all binaries have (non zero) orbital
motion. Hence, we should in principle include such motion in the
fit. We therefore begin by adding two linearized orbital motion
parameters $ds/dt$ and $d\alpha/dt$, i.e., the instantaneous time
derivatives of the separation and orientation of the binary axis.
However, we find that these parameters are quite poorly constrained,
with both the best fit and the majority of the chain being in the
unphysical $\beta>1$ regime, with very modest improvement in
$\chi^2$.  Here $\beta$ is the ratio of the projected kinetic to
projected potential energy \citep{don09} given by
\begin{equation}\label{beta}
\beta=\left(\frac{\rm KE}{\rm PE}\right)_\bot = \frac{\kappa M_\sun
{\rm yr}^2}{8\pi^2}\frac{\pi_{\rm E}s^3\gamma^2}{\theta_{\rm
E}(\pi_{\rm E}+\pi_{\rm S}/\theta_{\rm E})^3}
\end{equation}
where $\pi_{\rm S}={\rm AU}/D_{\rm S}$ is the parallax of the source
star at the distance $D_{\rm S}$,
$\gamma^2=(ds/dt/s)^2+(d\alpha/dt)^2$, $\kappa=(4G/c^2 {\rm
AU})\simeq 8.14\,{\rm mas\,yr}^{-1}$, and $\pi_{\rm
E}=\sqrt{\pi_{{\rm{E}},N}^{2}+\pi_{{\rm{E}},E}^{2}}$. We discuss the
evaluation of $\theta_{\rm E}$ in Section 4.2.

Since we know a priori that $\beta<1$, there is no information at
all in the very mild preference of the models for large $\beta$.
Under these circumstances, one might well consider suppressing
orbital motion by imposing $ds/dt = d\alpha/dt=0$, i.e., simply not
including these parameters in the fit.

However, in reality, all we are rigorously permitted to do by our
prior knowledge is to impose the physical constraint $\beta<1$. In
some circumstances, this would be effectively the same as ignoring
orbital motion altogether (i.e., $ds/dt = d\alpha/dt=0$). However,
as pointed out by \citet{bat11} and \citet{han16}, orbital-motion
parameters can be correlated with parallax parameters. Hence, they
must be included with the $\beta<1$ constraint.

In fact we impose a constraint $\beta<0.8$ because larger values of
$\beta$ require very unlikely physical situations observed at very
special angles. We find that the parallax parameters are in this
case essentially uncorrelated with the orbital motion parameters
thus restricted, so that the basic result derived by suppressing
orbital motion remains. However, the introduction of orbital motion
parameters does slightly increase the uncertainty in the parallax.
Hence, in our tables, we report results derived by including orbital
motion parameters but restricted to $\beta<0.8$. We report the
orbital motion parameters themselves in brackets because these
values are not measurements but rather simply reflect the prior
constraints.

\section{PHYSICAL QUANTITIES}
\subsection{Overview}\label{sec:phys-overview}
The total mass of lens system and its distance can be determined by
\begin{equation}\label{mass}
M_{\rm tot}=\frac{\theta_{\rm E}}{\kappa\pi_{\rm E}};\;D_{\rm
L}=\frac{{\rm AU}}{\pi_{\rm E}\theta_{\rm E}+\pi_{\rm S}}.
\end{equation}
If one applies Equation~(\ref{mass}) to the chain results summarized
in Table 1, the uncertainty in $M_{\rm tot}$ would appear to be
immense. First, since $\pi_{\rm E}$ is basically consistent with
zero, the fractional error in $\pi_{\rm E}$ (and so $M_{\rm tot}$)
should be almost infinite. In addition, for a microlensing event
with a resolved caustic crossing, $\theta_{\rm E}$ is unusually
poorly constrained. In the final analysis, this is due to the fact
that the source brightness $I_s$ is very poorly constrained. As in
the case of MOA-2011-BLG-293 \citep{yee12}, the large uncertainty in
$I_s$ is due to the extreme faintness of the source. That is, for
Einstein ring source positions $u\ll 1$, the solutions scale almost
perfectly in $(f_s,\rho,q,u_0, 1/t_{\rm E})$. (Here $f_s$ is the
OGLE source flux normalized at $I=18$.) Hence, almost all the
information on breaking this degeneracy comes from portions of the
light curve $u\sim 1$. However, because the source is extremely
faint, the fractional errors in the magnified source flux are high.
In addition, because this is a relatively low-cadence field for both
OGLE and KMTNet, there are relatively few data points in this part
of the light curve. If the source color is well-determined, then the
source size $\theta_* = K\sqrt{f_s}$ where $K$ is a constant that we
evaluate below. Then, $\theta_{\rm E}=\theta_*/\rho = K(f_s t_{\rm
E})/(t_*\sqrt{f_s})$. Since (from Table 1) the parameter
combinations $(f_s t_{\rm E})$ and $t_* \equiv \rho t_{\rm E}$ are
relatively well constrained, this means that $\theta_{\rm E}\propto
f_s^{-1/2}$. See also \citet{yee12}.

Nevertheless, there are two major pieces of auxiliary information
that can constrain the physical nature of the system in conjunction
with the microlens fit. First, there is a hard limit on the flux
from the lens, which (together with assumption that the host is
neither a neutron star nor a black hole), leads to a limit on the
lens mass. Second, the lens and source are both drawn from Galactic
populations whose statistical properties are known. In order to
assess the individual impact of these two pieces of ``external''
information, we introduce them sequentially. However, first we
evaluate $\theta_*$, or rather the constant $K$ that relates
$\theta_*$ to the source flux $f_s$ of an individual solution.

\subsection{Evaluation of $\theta_*$}\label{sec:phys-thetastar}
Because (from Table 1) $I_s$ is very poorly determined, we begin by
evaluating $\theta_*$ for an arbitrarily chosen fiducial value
\begin{equation}
I_{s,\rm fid} = 22.50 \label{eqn:fiducial}
\end{equation}
on the scale of the OGLE-IV observations. (As we discuss below, this
fiducial value is much brighter than the best fit models in Table
1). We measure the centroid of the OGLE-IV clump to be at $I_{\rm
clump} = $16.86. We adopt $I_{0,\rm clump} = $14.35 from
\citet{nat13} and thus derive
\begin{equation}
I_{s,0} = I_s + I_{0,\rm clump}-I_{\rm clump} = I_s - 2.51,
\label{eqn:Is0}
\end{equation}
and hence $I_{0,s,\rm fid} = 19.99$.

Similarly, we find the dereddened source color in $(V-I)$ from its
measured offset from the clump and the \citet{ben13} evaluation
$(V-I)_{0,\rm clump} = 1.06$. Unfortunately there are no magnified
$V$ data from OGLE and only one significantly magnified $V$ point
from KMT Chile (KMTC). To measure the source color we apply PyDIA
\citep{alb99} to the KMTC data, which yields DIA light curves and
photometry of neighboring stars on the same system\footnote{We note
that while DoPhot \citep{sch93} has been demonstrated by
\citet{ben13} to work well for sources that become bright $(I<16)$
during the event, it may work less well for faint sources.}.  We
show in Figure 5, a calibrated version of the PyDIA color-magnitude
diagram (CMD), which we obtain by aligning to OGLE-III
\citep{szy11}. Using PyDIA reductions of these data, we find an
offset $(V-I)_s - (V-I)_{\rm clump} = 0.29\pm 0.10$, and hence
\begin{equation}
(V-I)_{s,0} = (V-I)_s + (V-I)_{0,\rm clump}-(V-I)_{\rm clump} =
1.35\pm 0.10 \label{eqn:vmis0}
\end{equation}

Finally, we apply the $VIK$ color-color relations of \citet{bes88}
and the color/surface-brightness relations of \citet{ker04} to
obtain
\begin{equation}
\theta_* = (0.605\pm 0.024)\mu{\rm as} \times
10^{-0.2(I_{0,s}-I_{0,s,\rm fid})} = (0.605\pm 0.024)\mu{\rm as}
\times 10^{-0.2(I_s-22.50)} \label{eqn:thetastar-fid}
\end{equation}
We note that a typical main-sequence star with $(V-I)_{s,0}=1.35$,
lying at the distance of the bar toward this direction (7.4 kpc),
would have a dereddened magnitude $I_0\sim 20.4$, a point to which
we will return.

\subsection{Constraint on lens light}\label{sec:phys-lenslight}
The ``baseline object'' observed by OGLE has $I_{\rm base}=20.53$,
which $I_{\rm base}-I_{\rm clump}=3.98$ mag below the clump. This is
the brightness of a bulge turnoff star. Since this baseline object
must contain all the light from the source and lens (as well as
other possible ambient stars), we conservatively conclude that the
lens mass is $M<1.2\,M_\sun$.

We then impose this constraint on the same chain that was summarized
in Table 1. For each line in the chain, we take note of the values
of $(I_s,\rho,\pi_{{\rm E},N},\pi_{{\rm E},E})$. We evaluate
$\theta_*$ by inserting $I_s$ into
Equation~(\ref{eqn:thetastar-fid}) and then evaluate $\theta_{\rm E}
= \theta_*/\rho$ and $\pi_{\rm E} = (\pi_{{\rm E},N}^2+\pi_{{\rm
E},E}^2)^{1/2}$. We insert $\theta_{\rm E}$ and $\pi_{\rm E}$ into
Equation~(\ref{mass}) and reject the link if the resulting value
violates the mass constraint. Figure 6 presents the distribution of
parallax parameters after the mass constraint, and Table 2 shows the
resulting microlensing parameters.

Comparison of these results with Table 1 reveals several significant
differences. Before commenting on these, however, it is important to
note that the ``conserved quantities'' below the line actually
change very little. As derived in Section 4.1, the fact that $(f_s
t_{\rm E})$ and $t_*=(\rho t_{\rm E})$ are nearly invariant, implies
that $\theta_{\rm E} = K(f_s t_{\rm E})/(t_*\sqrt{f_s})$ basically
scales $\propto f_s^{-1/2}$. Hence, it can be well approximated by
\begin{equation}
M_{\rm tot} \propto f_s^{-1/2}\pi_{\rm E}^{-1}. \label{eqn:mscale}
\end{equation}
From the first term, it follows that imposing a mass limit will tend
to suppress solutions with small $f_s$ and therefore, from the
approximate invariance of $(f_s t_{\rm E})$, $(q t_{\rm E})$, $t_*$,
and $t_{\rm eff}$, also suppress solutions with large $t_{\rm E}$,
small $q$, small $\rho$ and small $u_0$. The differences in these
parameters between Tables 1 and 2 is uniformly a factor 2.2 for
$s<1$ and 2.6 for $s>1$. We note that although the proper motion
$\mu=\theta_*/t_*$ is not explicitly shown in these tables, it is
directly affected, $\mu=\theta_*/t_* \propto f_s^{1/2}$, i.e.,
pushed upward.

The second term in Equation~(\ref{eqn:mscale}) directly suppresses
solutions with small $\pi_{\rm E}$. Because (as is often the case)
$\pi_{{\rm E},E}$ is much better constrained by the light curve than
$\pi_{{\rm E},N}$, the main impact of the mass constraint is on
$\pi_{{\rm E},N}$ which then becomes approximately bimodal,
especially for the $s<1$ solution. To better capture this behavior,
we show in Table 2 the absolute value of $\pi_{{\rm E},N}$, rather
than the parameter itself.

\subsection{Galactic Model Prior}\label{sec:phys-galprior}
The Galactic model prior is evaluated using the method of
\citet{bat11}, and the prior distribution of the Galactic model is
given by
\begin{equation}\label{prior1}
P_{\rm Gal}\propto2R_{\rm E}v_{\rm
rel}\nu(x,y,z)f(\mbox{\boldmath$\mu$})g(M),
\end{equation}
where $R_{\rm E}$ is the physical Einstein ring radius, $v_{\rm
rel}=\mu D_{\rm L}$ is the lens velocity relative to the
observer-source line of sight, $\mbox{\boldmath$\mu$}$ is the
lens-source relative proper motion, $\nu(x,y,z)$ is the density
distribution of the lenses. We adopt the \citet{han03} model which
includes the bulge model of \citet{dwe95} and disk model of
\citet{zhe01}. Here $g(M)$ is the mass function, for which we adopt
$g(M)\propto M^{-1}$ in common with \citet{bat11}, and
$f(\mbox{\boldmath$\mu$})$ is the probability function for the
lens-source relative proper motion. Equation~(\ref{prior1}) can be
expressed for microlensing parameters as
\begin{equation}\label{prior2}
P_{\rm Gal}\propto \frac{4D_{\rm L}^4\mu^4}{{\rm AU}\pi_{\rm
E}}\nu(x,y,z)f(\mbox{\boldmath$\mu$})
\end{equation}
We apply the Galactic model priors to each of the chain links. We
show microlensing parameters after imposing these priors in Table 3.

By comparing Table 3 with Table 2, we can understand the impact of
these priors. The main impact comes from the Jacobian terms that are
needed to transform Equation (10) to Equation (11). That is, as
discussed by \citet{bat11} the microlensing chains have a uniform
prior in $\mbox{\boldmath$\pi$}_{\rm E}$, for which phase space is
dominated by short $D_L$, whereas for Euclidean geometry, phase
space is dominated by large distances. This is origin of the terms
``$D_{\rm L}^4\mu^4$''. As discussed in
Section~\ref{sec:phys-lenslight}, $\mu\propto f_s^{1/2}$. Hence, the
$\mu^4$ term strongly favors higher $f_s$ and therefore also the
corresponding changes in $(t_{\rm E},q,\rho,u_0)$ implied by the
invariant quantities at the bottom of Tables 1--3.

The $D_{\rm L}^4$ term works in the same direction, albeit more
weakly. As shown in Section~\ref{sec:phys-overview}, $\theta_{\rm
E}\propto f_s^{-1/2}$. Hence, high $f_s$ leads to lower $\theta_{\rm
E}$, lower $\pi_{\rm rel} \equiv \theta_{\rm E}\pi_{\rm E}$, and so
(via Equation (4)) higher $D_{\rm L}$. However, this effect is
somewhat ``watered down'' by the appearance of $\pi_{\rm S}$ is the
denominator. The $D_{\rm L}$ term similarly favors solutions with
smaller $\pi_{\rm E}$, as of course does the direct appearance of
$\pi_{\rm E}$ in the denominator of Equation (11).

All of these effects are seen in the comparison of Table 3 with
Table 2. A more detailed analysis would show that the
``$\nu(x,y,z)$'' and ``$f(\mbox{\boldmath$\mu$})$'' terms act in the
same direction, but more weakly. Comparing Table 3 to Table 1, one
sees that the lens-light constraint and the Galactic priors combine
to greatly reduce the range of microlensing parameters.

\subsection{Transformation from Microlensing to Physical Parameters}
\label{sec:phys-phys} In Figure 7, we present the histograms for the
host $M_{\rm 1}$ and companion $M_{\rm 2}$ mass, the projected
separation between them $a_{\bot}$, and the distance of the lens
system $D_{\rm L}$. The histograms with black and red lines show the
distributions of physical quantities from MCMC runs and the Galactic
model prior including host-mass cut, respectively. In Table 4, we
present the physical quantities with $1\,\sigma$ error for two
parallax+orbital motion models and applying the Galactic model prior
(including host mass cut). In the case of the physical quantities
obtained by applying the Galactic model prior, we mark them in
brackets. The main effect of applying the priors is to shift the
masses of the brown dwarf and host sharply higher and also to place
the system at significantly larger distance. As discussed in
Section~\ref{sec:phys-galprior}, the primary reason for this is not
the Galactic model parameters, but the Jacobian term, which in turn
overwhelmingly reflects the ratio of phase space available to chain
relative to the real physical phase space available to the lens.

In the present case, the mass cut (derived from the lens flux
constraint) also plays a significant role. This is because the
parallax measurement is consistent with zero, which means that the
Jacobian itself tends to be unrestricted in its ``preference'' for
high mass solutions. However, just as the microlens parallax cannot
actually be zero, it also cannot be so small as to imply a huge lens
mass, even though this is not ruled out by the MCMC chain itself.

Based on the results summarized in Table 4 and Figure 7,
OGLE-2016-BLG-0693LB is well constrained to be a brown dwarf in
either the close $(s<1)$ or wide $(s>1)$ solutions, which are fully
degenerate (see first row of Table 1). Its host is almost certainly
an FGK star, i.e., the traditional targets of RV surveys on which
the brown dwarf desert was originally defined by \citet{mar00},
although there is a small probability that the host is an early M
dwarf.  The original definition of the ``brown dwarf desert''
$(a<3\,{\rm AU})$ was set by observational limits of RV surveys at
that time.  In the close solution, OGLE-2016-BLG-0693LB is within
the brown dwarf desert according to this restrictive definition. In
the wide solution, it lies at or somewhat beyond its outer edge.
This detection therefore illustrates the prospects for microlensing
to probe the outer edges of the brown dwarf desert, where it is
sensitive to both brown dwarfs and massive planets, and can thus
determine whether the brown dwarfs remain a minimum in the companion
mass function as they are at closer separations.

\subsection{Two Consistency Checks}\label{sec:phys-consist}
The solutions derived in Tables 3 and 4 did not depend in any way on
the source magnitude being ``reasonable'' given its color and only
very weak prior information on the lens-source relative proper
motion $\mbox{\boldmath$\mu$}$. That is, as we noted in
Section~\ref{sec:phys-galprior}, the Galactic prior does have a
strong $\mu$ dependence, but this is overwhelmingly due to the
Jacobian term $\mu^4$ and is only weakly responsive to the
Galactic-model assumption $f(\mbox{\boldmath$\mu$})$. Therefore, we
can ask whether the final values of these quantities are consistent
with what is expected given our knowledge of the Galaxy, as external
checks on the solution.

The best-fit source magnitudes in the $s<1$ and $s>1$ solutions are
$I_s=22.3\pm 0.4$ and $I_s=22.5\pm 0.3$, respectively. Recall from
Section~\ref{sec:phys-thetastar} that the expected value for a
typical star of the source's color and lying in Galactic bar would
be $I_s=22.9$. Since variations in depth within the bar and
source-star metallicity could cause this to vary by $\pm 0.3$ mag,
either of these models is consistent with expectations at the
$1\,\sigma$ level.

The model proper motions ($\mu=1.7\pm 0.4$ and $\mu=1.4\pm 0.2$) at
first sight appear more problematic given that the lens is in the
disk and source is (at least assumed to be) in the bulge. This is
because, if we assume ignore the peculiar motions of the observer,
lens, and source, and if we assume a flat rotation curve and also
assume zero bulk motion for the bulge, then the expected proper
motion is just $\mu = v_{\rm rot}/D_{\rm S} = 6.3\,{\rm mas}\,{\rm
yr}^{-1}$, where we have adopted $v_{\rm rot} = 220\,\,{\rm
km}\,{\rm s}^{-1}$ and $D_S=7.4$ kpc. Then it would appear to
require some fine-tuning to ``arrange'' for the peculiar motions to
almost perfectly cancel out this mean motion in order to yield
$\mu\sim 1.5\,{\rm mas}\,{\rm yr}^{-1}$.

However, the Galactic longitude of this event is unusually high,
$l=5.5^\circ$. The Galactic bar is certainly rotating. The exact
angular speed is not precisely known, but we can plausibly adopt
$100\,{\rm km}\,{\rm s}^{-1}\,{\rm kpc}^{-1}$. Then the mean motion
projected on the plane of the sky at the source position is $\sim
100\,{\rm km}\,{\rm s}^{-1}$. For the geometry of this event (and
now taking account of the Solar peculiar motion) the mean proper
motion is $\overline{\mbox{\boldmath$\mu$}} = (3.0,0.0)\,{\rm
mas}\,{\rm yr}^{-1}$ in Galactic $(l,b)$ coordinates. Note that in
Equatorial coordinates, this direction is about $30^\circ$ East of
North, which is very similar to the direction of
$\mbox{\boldmath$\pi$}_{\rm E}$ (i.e., the same as
\mbox{\boldmath$\mu$}). Given this mean motion, the observed proper
motion is not at all unexpected.

Finally, we recall that $t_* \simeq 3.6\,$hr is a near-invariant,
i.e., it is a nearly model-independent property of the light curve.
This long duration was actually essential to the publication of this
paper. Recall that the event lies in a field with KMT cadence
$\Gamma=0.4\,{\rm hr}^{-1}$. It was only this long $t_*$ that
permitted accurate measurement of the caustic crossing time. Even
with the long $t_*$, there were only two data points over the
caustic, which is the minimum for a reliable measurement. However,
given that the great majority of microlensing sources are
main-sequence stars (include the source star in the present case),
such a long $t_* = \theta_*/\mu$ is only possible if the proper
motion is small.

\section{CONCLUSION}
The event of OGLE-2016-BLG-0693 was discovered based on the
survey-only microlensing observations of the OGLE and KMTNet groups.
Although the event lies in a relatively low cadence KMTNet field
($0.4$ ${\rm hr}^{-1}$), the caustic crossing part was observed by
KMTS. This became possible because the source crossing time was
exceptionally long (given that the source is a dwarf), $t_*=3.6$ hr,
which permitted two flux measurements over the caustic crossing. The
long crossing time was in turn due to a proper motion, $\mu=
1.55\pm0.3$ ${\rm mas\,yr^{-1}}$, that is much lower than typical
for most microlensing events, but not unexpected for events seen at
high positive Galactic longitude ($5.5^\circ$) on the near side of
the Galactic bar.

In the analysis of the event, we considered the effect of parallax
according to Earth's orbital motion during the lensing phenomenon
because the event has an unusually long timescale (which is another
consequence of its unusually low proper motion). In addition, we
took the effects of baseline slope into account. We conducted the
analysis of the light curve for two microlensing models with close
and wide separation, respectively.

After applying a Galactic model prior and a lens-mass upper limit
(due to limits on lens light), we found that the lens system is
composed of a low-mass brown dwarf ($49^{+20}_{-18}\,M_J$) orbiting
a K or G dwarf. Finally, we note that the separation of the brown
dwarf is less than $\sim 5$ AU demonstrating that microlensing is
poised to probe the brown dwarf desert at intermediate separations.

\acknowledgements We thank the anonymous referee for a very helpful
report. This research has made use of the KMTNet system operated by
the Korea Astronomy and Space Science Institute (KASI) and the data
were obtained at three host sites of CTIO in Chile, SAAO in South
Africa, and SSO in Australia. OGLE project has received funding from
the National Science Centre, Poland, grant MAESTRO
2014/14/A/ST9/00121 to AU. AG, YKJ, and WZ were supported by US NSF
grant AST-1516842. Work by IGS was supported by JPL grant 1500811.
Work by C.H. was supported by the grant (2017R1A4A1015178) of
National Research Foundation of Korea. YHR was supported by KASI
(Korea Astronomy and Space Science Institute) grant 2017-1-830-03.

\clearpage

\begin{figure}
\plotone{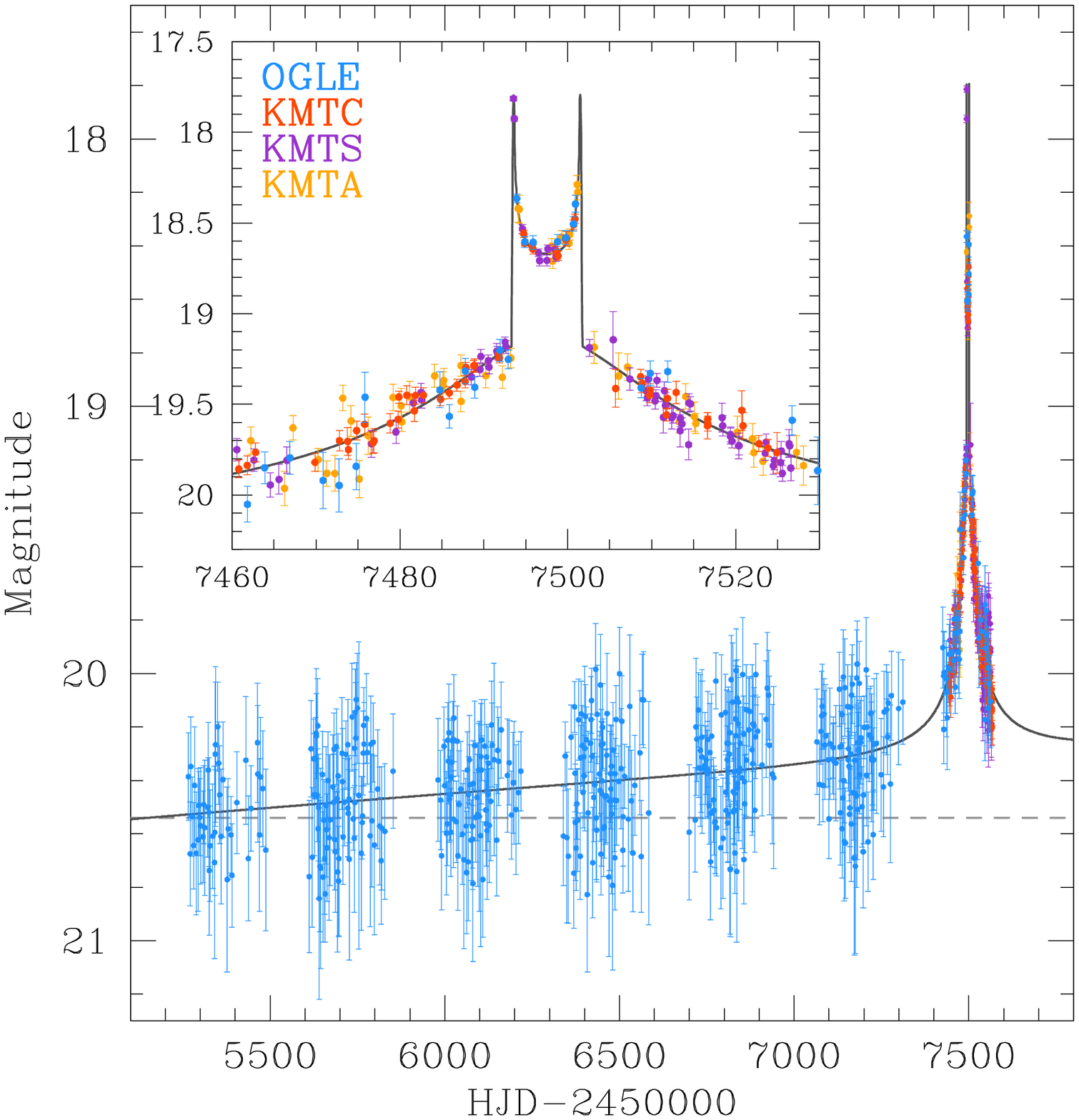} \caption{Full light curve data from OGLE (blue) and
three KMTNet sets from KMTC (Chile, red), KMTS (South Africa,
violet), and KMTA (Australia, yellow). Inset shows the data near the
caustic crossing. Model light curve with black line includes the
effect of baseline slope.}
\end{figure}

\begin{figure}
\plotone{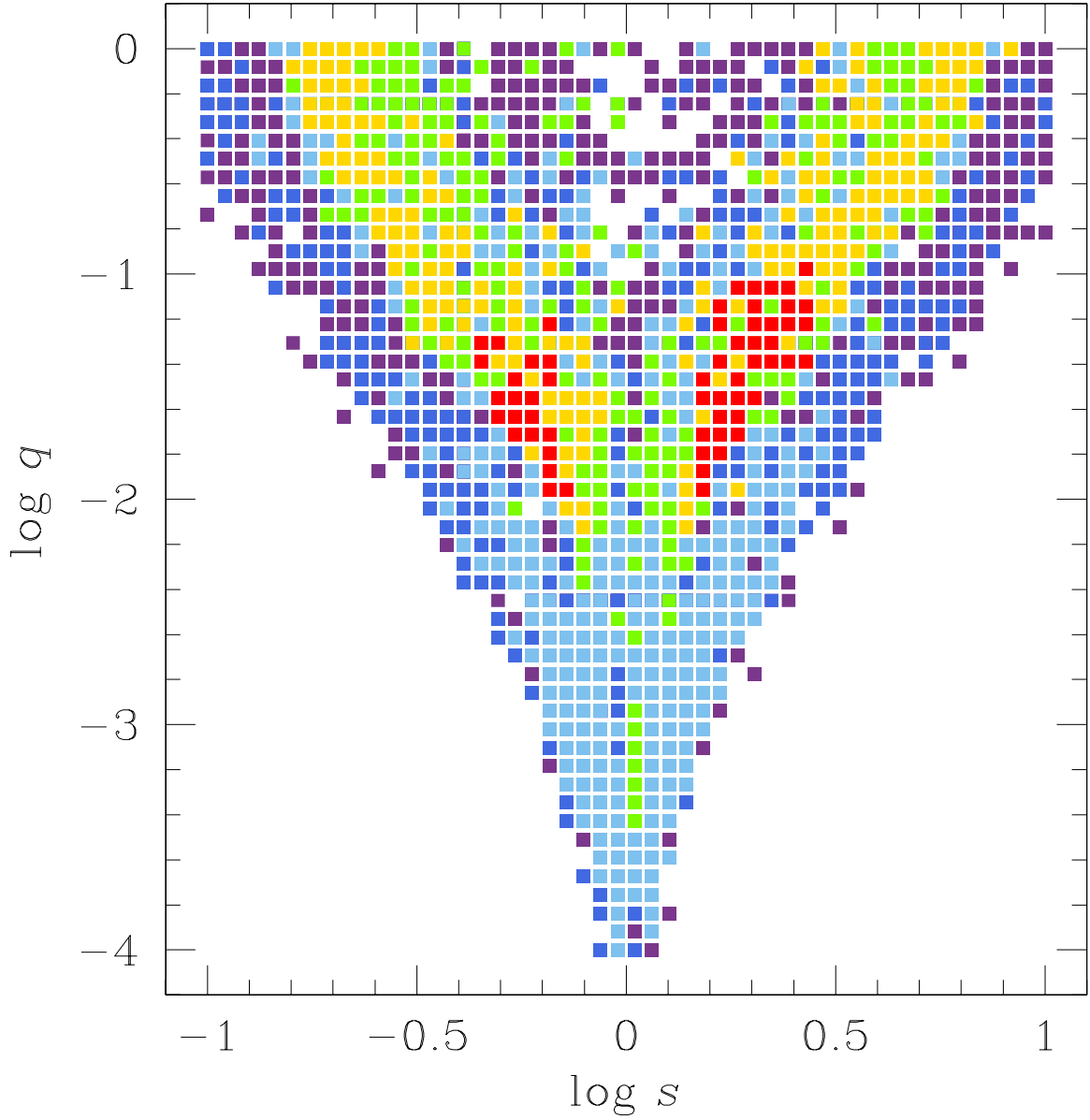} \caption{The $\Delta\chi^2$ map for $(s,q)$
parameters. The $\Delta\chi^2$ with $<100,<400,<900,<1600,<2500,$
and $<3600$ levels presented as red, yellow, green, light blue,
blue, and purple, respectively.}
\end{figure}

\begin{figure}
\plotone{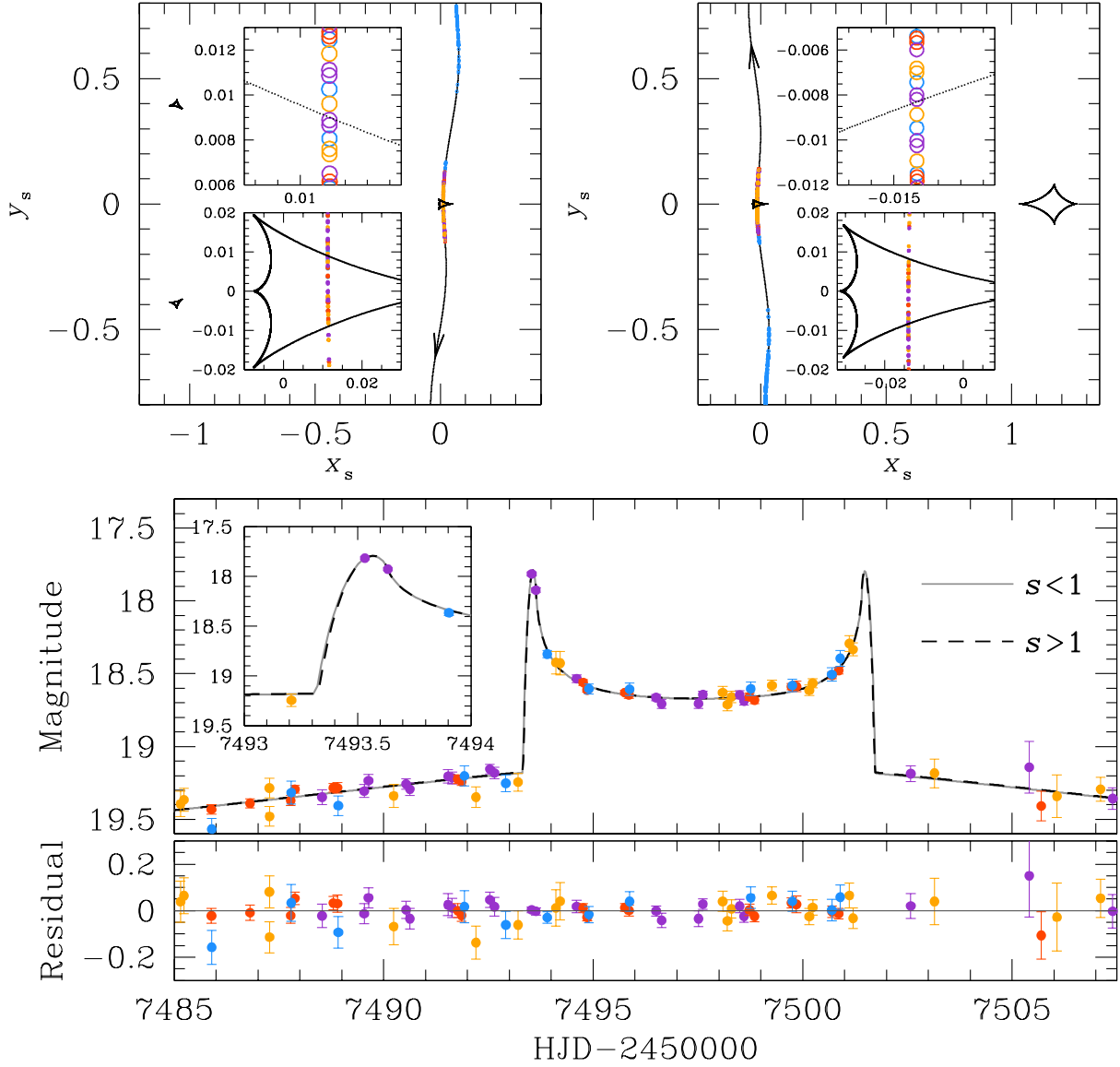} \caption{Caustic and light curve of the parallax
models with the effect of slope. The upper two panels represent the
caustic structure for the close- (left-hand panel) and
wide-separation (right-hand panel) models. The black solid line with
an arrow is the source trajectory, and the circles with various
colors show the source positions corresponding to the time of
observation of each observatory. The lower panel shows the light
curves of the caustic crossing parts for close- and wide-separation
models.}
\end{figure}

\begin{deluxetable}{lccccc}
\tablecolumns{6} \tablewidth{0pc}\tablecaption{\textsc{Best-fit
solutions}} \tablehead{ \colhead{ } &
\multicolumn{2}{c}{Parallax models}& \colhead{ }&\multicolumn{2}{c}{Parallax+Orbital motion models}\\
\cline{2-3} \cline{5-6} \colhead{Parameters } &
\colhead{$s<1$}&\colhead{$s>1$}& \colhead{
}&\colhead{$s<1$}&\colhead{$s>1$}} \startdata
  $\chi^2/\rm{dof}$               &767.217/756           &767.304/756           & &766.769/754           &767.126/754          \\
  $t_0$ $(\rm{HJD}^{\prime})$     &7497.484 $\pm$ 0.087  &7497.559 $\pm$ 0.158  & &7497.465 $\pm$ 0.125  &7497.561 $\pm$ 0.168 \\
  $u_0$                           &0.011 $\pm$ 0.006     &0.014 $\pm$ 0.010     & &0.010 $\pm$ 0.010     &0.014 $\pm$ 0.011    \\
  $t_{\rm E}$ $(\rm{days})$       &451.190 $\pm$ 136.631 &488.108 $\pm$ 287.923 & &507.369 $\pm$ 127.955 &556.115 $\pm$ 128.564\\
  $s$                             &0.592 $\pm$ 0.030     &1.760 $\pm$ 0.044     & &0.582 $\pm$ 0.039     &1.788 $\pm$ 0.044    \\
  $q$ $(10^{-2})$                 &2.127 $\pm$ 0.805     &2.077 $\pm$ 1.532     & &1.931 $\pm$ 1.261     &1.902 $\pm$ 1.584    \\
  $\alpha$ $(\rm{rad})$           &1.562 $\pm$ 0.019     &4.718 $\pm$ 0.022     & &1.581 $\pm$ 0.041     &4.719 $\pm$ 0.022    \\
  $\rho$ $(10^{-4})$              &3.296 $\pm$ 1.462     &2.987 $\pm$ 2.287     & &2.872 $\pm$ 2.726     &2.689 $\pm$ 2.426    \\
  $\pi_{\rm{E},\it{N}}$           &-0.017 $\pm$ 0.125    &0.014 $\pm$ 0.099     & &0.010 $\pm$ 0.162     &0.003 $\pm$ 0.141    \\
  $\pi_{\rm{E},\it{E}}$           &0.011 $\pm$ 0.020     &0.012 $\pm$ 0.019     & &0.011 $\pm$ 0.027     &0.011 $\pm$ 0.025    \\
  $\rm{Slope}$ $(\rm{yr}^{-1})$   &0.038 $\pm$ 0.007     &0.038 $\pm$ 0.008     & &0.037 $\pm$ 0.009     &0.036 $\pm$ 0.008    \\
  $ds/dt$ $(\rm{yr}^{-1})$        &-                     &-                     & &[0.259 $\pm$ 0.628]   &[-0.107 $\pm$ 0.471]    \\
  $d\alpha/dt$ $(\rm{yr}^{-1})$   &-                     &-                     & &[0.039 $\pm$ 0.316]   &[0.091 $\pm$ 0.277]    \\
  $I_s$                           &23.649 $\pm$ 0.378    &23.731 $\pm$ 0.584    & &23.812 $\pm$ 0.476    &23.882 $\pm$ 0.433     \\
  $I_b$                           &20.587 $\pm$ 0.039    &20.584 $\pm$ 0.057    & &20.577 $\pm$ 0.068    &20.573 $\pm$ 0.057     \\
  \cline{1-6}
  $t_*$ $(\rm{days})$             &0.148 $\pm$ 0.021     &0.146 $\pm$ 0.023     & &0.146 $\pm$ 0.025     &0.149 $\pm$ 0.021     \\
  $t_{\rm eff}$ $(\rm{days})$     &5.122 $\pm$ 0.788     &6.812 $\pm$ 1.608     & &4.817 $\pm$ 0.809     &7.802 $\pm$ 1.639     \\
  $qt_{\rm E}$ $(\rm{days})$      &9.597 $\pm$ 1.187     &10.140 $\pm$ 0.936    & &9.799 $\pm$ 1.338     &10.579 $\pm$ 0.935     \\
  $f_{\rm s}t_{\rm E}$            &2.483 $\pm$ 0.213     &2.491 $\pm$ 0.189     & &2.402 $\pm$ 0.218     &2.469 $\pm$ 0.184     \\
\enddata
\tablecomments{$\textrm{HJD}^\prime=\textrm{HJD}-2450000$. $ds/dt$
and $d\alpha/dt$ cannot be regarded as ``measured": They basically
reflect the physically motivated condition $\beta<0.8$.}
\end{deluxetable}

\begin{figure}
\plotone{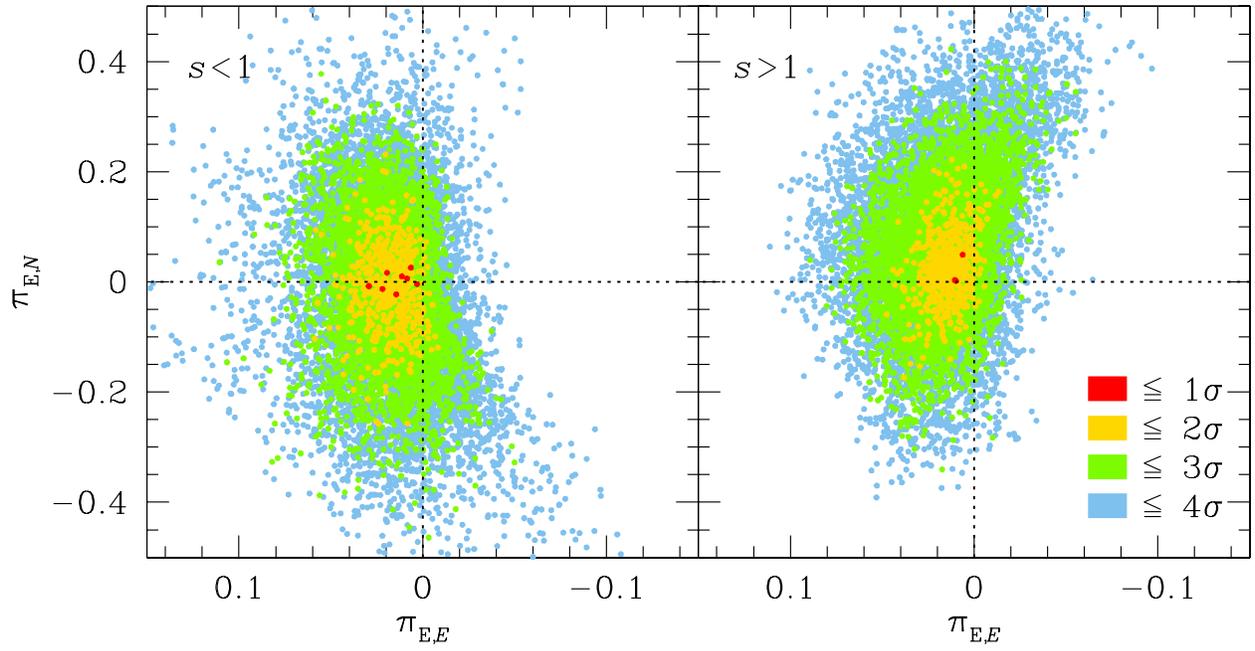} \caption{Contours for parallax parameters derived
from MCMC fits. The colors of red, yellow, green, light blue
correspond to $\Delta\chi^2<1,<4,<9$, and $<16$, respectively.}
\end{figure}

\begin{figure}
\plotone{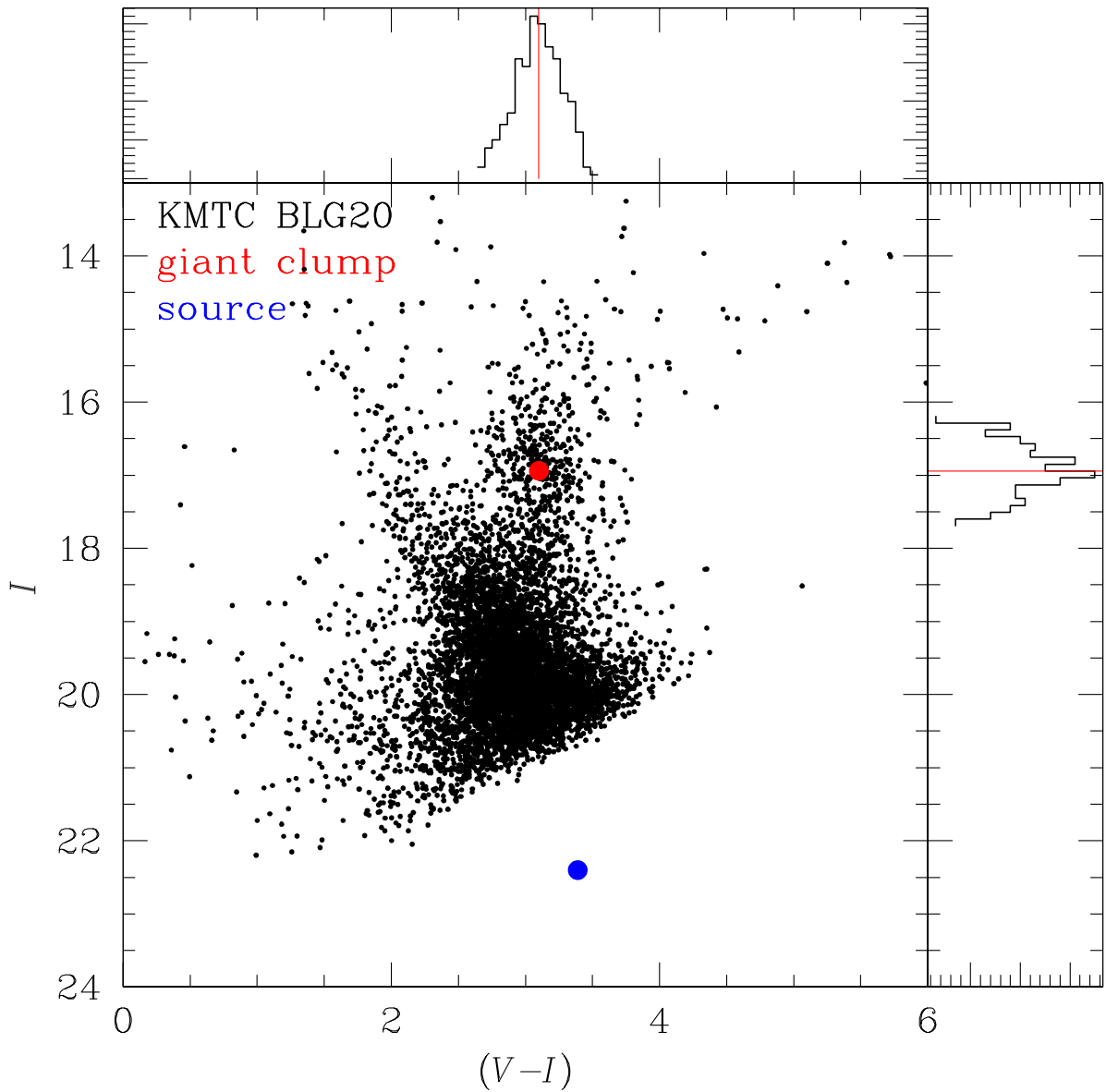} \caption{Color-magnitude diagram (CMD) of field
stars around the source star in KMTNet CTIO BLG20 field. The CMD
shows the microlensed source (blue) and centroid of the giant clump
(red). The upper and right-side panels show the column density of
stars around the giant clump. The algorithm centers on the 2-D peak
and the histograms show 1-D distributions, which indicates good
agreement. Note that the magnitude of the source flux is for the
best fit, but the errors in this quantity are very large. See text
for how these large error bars were handled.}
\end{figure}

\begin{figure}
\plotone{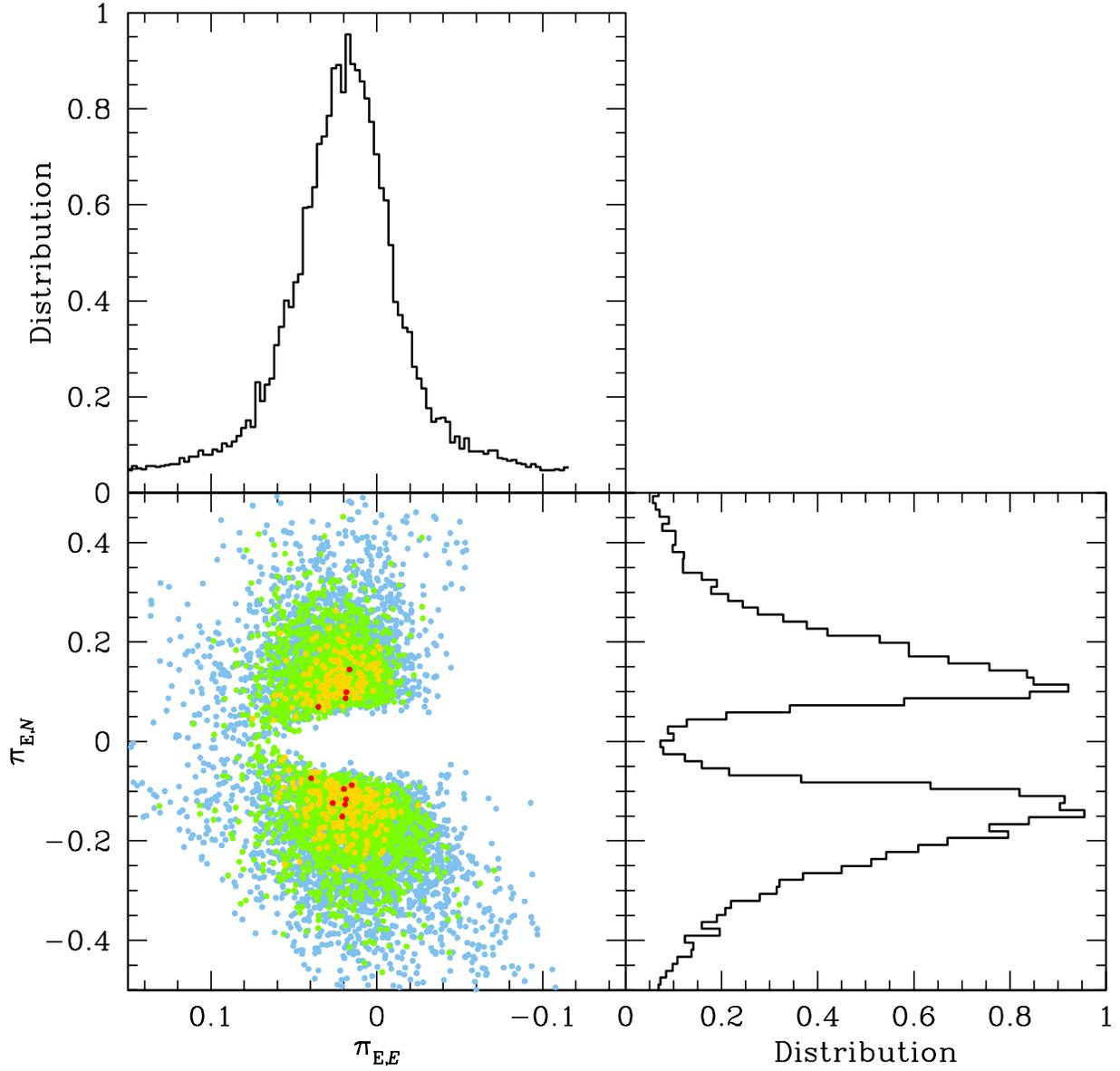} \caption{Contours and histograms for parallax
parameters after the mass constraint.}
\end{figure}

\begin{deluxetable}{lcc}
\tablecolumns{3} \tablewidth{0pc}\tablecaption{\textsc{Lensing
parameters after the mass constraint}}
\tablehead{\colhead{ } &\multicolumn{2}{c}{Parallax+Orbital motion models}\\
\cline{2-3} \colhead{Parameters } &\colhead{$s<1$}&\colhead{$s>1$}}
\startdata
  $t_0$ $(\rm{HJD}^{\prime})$     &7497.440 $\pm$ 0.134  &7497.616 $\pm$ 0.168 \\
  $u_0$                           &0.022 $\pm$ 0.011     &0.029 $\pm$ 0.011    \\
  $t_{\rm E}$ $(\rm{days})$       &244.540 $\pm$ 103.180 &229.050 $\pm$ 83.785\\
  $s$                             &0.628 $\pm$ 0.040     &1.754 $\pm$ 0.047    \\
  $q$ $(10^{-2})$                 &3.705 $\pm$ 1.321     &4.754 $\pm$ 1.542    \\
  $\alpha$ $(\rm{rad})$           &1.598 $\pm$ 0.044     &4.736 $\pm$ 0.023    \\
  $\rho$ $(10^{-4})$              &6.101 $\pm$ 2.828     &6.892 $\pm$ 2.357    \\
  $|\pi_{\rm{E},\it{N}}|$         &0.165 $\pm$ 0.088     &0.159 $\pm$ 0.091    \\
  $\pi_{\rm{E},\it{E}}$           &0.018 $\pm$ 0.029     &0.016 $\pm$ 0.029    \\
  $\rm{Slope}$ $(\rm{yr}^{-1})$   &0.047 $\pm$ 0.008     &0.047 $\pm$ 0.007    \\
  $ds/dt$ $(\rm{yr}^{-1})$        &[0.589 $\pm$ 0.712]   &[-0.672 $\pm$ 0.498]    \\
  $d\alpha/dt$ $(\rm{yr}^{-1})$   &[-0.522 $\pm$ 0.307]  &[-0.228 $\pm$ 0.303]    \\
  $I_s$                           &22.949 $\pm$ 0.469    &22.841 $\pm$ 0.368        \\
  $I_b$                           &20.660 $\pm$ 0.068    &20.675 $\pm$ 0.054       \\
\cline{1-3}
  $t_*$ $(\rm{days})$             &0.151 $\pm$ 0.024     &0.161 $\pm$ 0.020       \\
  $t_{\rm eff}$ $(\rm{days})$     &5.533 $\pm$ 0.989     &6.770 $\pm$ 1.738        \\
  $qt_{\rm E}$ $(\rm{days})$      &8.807 $\pm$ 1.326     &10.926 $\pm$ 0.956       \\
  $f_{\rm s}t_{\rm E}$            &2.616 $\pm$ 0.272     &2.686 $\pm$ 0.182        \\
\enddata
\end{deluxetable}

\begin{deluxetable}{lcc}
\tablecolumns{3} \tablewidth{0pc}\tablecaption{\textsc{Lensing
parameters weighted by the Galactic model prior}}
\tablehead{\colhead{ } &\multicolumn{2}{c}{Parallax+Orbital motion models}\\
\cline{2-3} \colhead{Parameters } &\colhead{$s<1$}&\colhead{$s>1$}}
\startdata
  $t_0$ $(\rm{HJD}^{\prime})$     &7497.451 $\pm$ 0.123  &7497.555 $\pm$ 0.156 \\
  $u_0$                           &0.039 $\pm$ 0.013     &0.037 $\pm$ 0.010    \\
  $t_{\rm E}$ $(\rm{days})$       &142.510 $\pm$ 50.025  &169.250 $\pm$ 42.235\\
  $s$                             &0.677 $\pm$ 0.039     &1.731 $\pm$ 0.055    \\
  $q$ $(10^{-2})$                 &5.174 $\pm$ 1.100     &6.229 $\pm$ 1.525    \\
  $\alpha$ $(\rm{rad})$           &1.607 $\pm$ 0.044     &4.736 $\pm$ 0.019    \\
  $\rho$ $(10^{-4})$              &9.838 $\pm$ 3.124     &8.959 $\pm$ 2.378    \\
  $|\pi_{\rm{E},\it{N}}|$         &0.087 $\pm$ 0.048     &0.078 $\pm$ 0.043    \\
  $\pi_{\rm{E},\it{E}}$           &0.029 $\pm$ 0.034     &0.032 $\pm$ 0.029    \\
  $\rm{Slope}$ $(\rm{yr}^{-1})$   &0.053 $\pm$ 0.010     &0.050 $\pm$ 0.008    \\
  $ds/dt$ $(\rm{yr}^{-1})$        &[0.661 $\pm$ 0.705]   &[-0.912 $\pm$ 0.346]    \\
  $d\alpha/dt$ $(\rm{yr}^{-1})$   &[-0.627 $\pm$ 0.258]  &[-0.288 $\pm$ 0.317]    \\
  $I_{\rm s}$                     &22.317 $\pm$ 0.372    &22.547 $\pm$ 0.289        \\
  $I_{\rm b}$                     &20.764 $\pm$ 0.086    &20.722 $\pm$ 0.059       \\
   \cline{1-3}
  $t_*$ $(\rm{days})$             &0.145 $\pm$ 0.024     &0.154 $\pm$ 0.020       \\
  $t_{\rm eff}$ $(\rm{days})$     &5.632 $\pm$ 0.518     &6.105 $\pm$ 1.915        \\
  $qt_{\rm E}$ $(\rm{days})$      &7.576 $\pm$ 1.109     &10.543 $\pm$ 1.000       \\
  $f_{\rm s}t_{\rm E}$            &2.661 $\pm$ 0.144     &2.602 $\pm$ 0.158        \\
\enddata

\end{deluxetable}

\begin{deluxetable}{lcc}
\tablecolumns{3} \tablewidth{0pc}\tablecaption{\textsc{Physical
properties of the binary system}}
\tablehead{\colhead{Quantity}&\colhead{$s<1$}&\colhead{$s>1$}}
\startdata
  $M_1$ $[M_\sun]$       &$0.59_{-0.28}^{+0.37}$      &$0.57_{-0.26}^{+0.37}$     \\
 (prior)                 &$(0.86_{-0.32}^{+0.24})$    &$(0.89_{-0.30}^{+0.22})$   \\
  $M_2$ $[M_J]$          &$21.39_{-8.20}^{+12.84}$    &$25.41_{-10.47}^{+18.42}$   \\
 (prior)                 &$(42.62_{-16.29}^{+18.22})$ &$(54.75_{-19.25}^{+21.45})$ \\
  $D_{\rm L}$ [kpc]      &$3.64_{-0.97}^{+0.99}$      &$3.82_{-0.97}^{+1.07}$     \\
 (prior)                 &$(4.96_{-1.02}^{+0.64})$    &$(5.19_{-0.87}^{+0.58})$   \\
  $a_\bot$ [AU]          &$1.85_{-0.48}^{+0.45}$      &$5.12_{-1.38}^{+1.25}$     \\
 (prior)                 &$(2.24_{-0.35}^{+0.32})$    &$(5.80_{-0.80}^{+0.64})$   \\
   $\mu$ [mas/yr]        &$1.21_{-0.28}^{+0.40}$      &$1.15_{-0.23}^{+0.24}$     \\
 (prior)                 &$(1.62_{-0.36}^{+0.45})$    &$(1.38_{-0.23}^{+0.28})$   \\
\enddata
\end{deluxetable}

\begin{figure}
\plotone{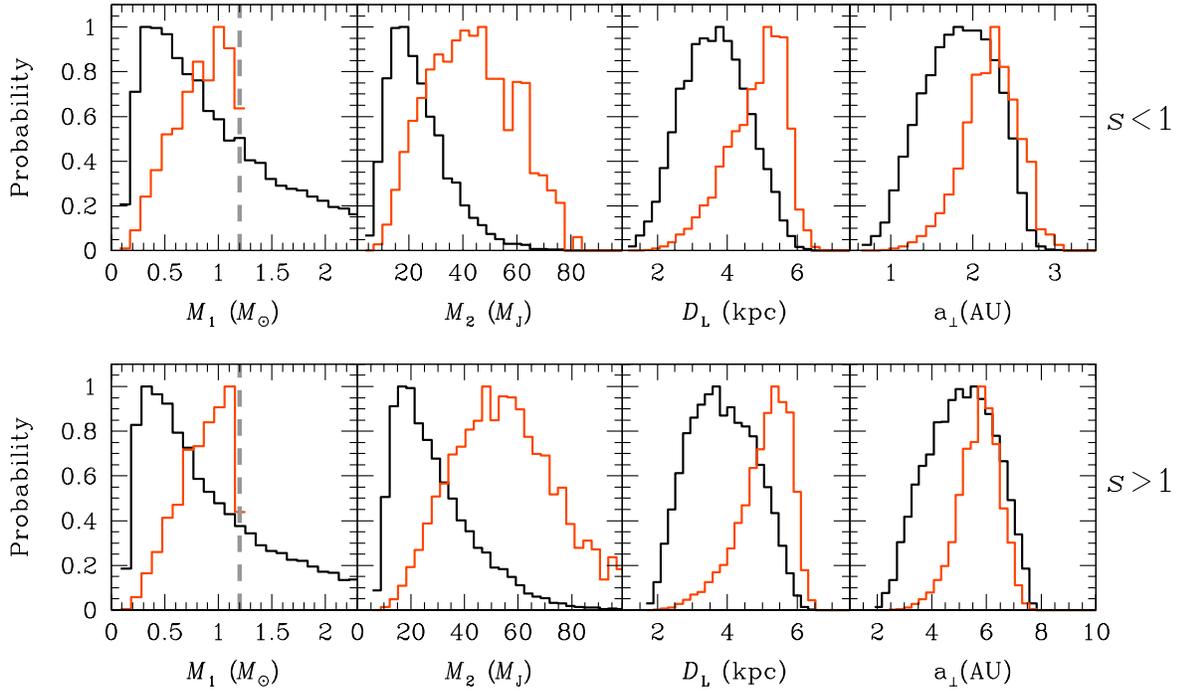} \caption{Histograms for physical quantities of the
lens. The histograms with the black line represent the uncertainties
of physical quantities from the MCMC runs, and the histograms with
red line show that after applying the Galactic model prior
(including the lens-mass cut $M<1.2\,M_\sun$; gray dashed line).
Large offsets between the black and red histograms is overwhelmingly
due to the Jacobian term, which is the ratio of phase space
available to the lens relative to that available to the MCMC chain.
See text.}
\end{figure}

\end{document}